\documentclass[english,twocolumn,showpacs,preprintnumbers,amsmath,amssymb,pre]{revtex4}
\usepackage{mathptmx}
\usepackage[T1]{fontenc}
\usepackage[latin9]{inputenc}
\usepackage{color}
\usepackage{amstext}
\usepackage{graphicx}
\usepackage{esint}
\usepackage{amsmath}

\makeatletter

\@ifundefined{textcolor}{}
{%
 \definecolor{BLACK}{gray}{0}
 \definecolor{WHITE}{gray}{1}
 \definecolor{RED}{rgb}{1,0,0}
 \definecolor{GREEN}{rgb}{0,1,0}
 \definecolor{BLUE}{rgb}{0,0,1}
 \definecolor{CYAN}{cmyk}{1,0,0,0}
 \definecolor{MAGENTA}{cmyk}{0,1,0,0}
 \definecolor{YELLOW}{cmyk}{0,0,1,0}
}

\makeatother

\usepackage{babel}
\begin{document}

\title{Topological phase transition in a discrete quasicrystal}

%\author{Eran Sagi \cite{current}}
%\author{Eli Eisenberg}

%\author{Eran Sagi$^{1,2}$ , Eli Eisenberg$^{1}$\\
%{$^1$\small \em Raymond and Beverly Sackler School of Physics and Astronomy, Tel
%Aviv University, Tel Aviv 69978, Israel}\\
%{$^2$\small \em Current address: Department of Condensed Matter Physics, Weizmann Institute of Science, Rehovot 76100, Israel}
%}
\author{Eran Sagi}
\altaffiliation[Current address: ]{Department of Condensed Matter Physics, Weizmann Institute of Science, Rehovot 76100, Israel.}
%\affiliation{TAU}
\author{Eli Eisenberg}
\affiliation{Raymond and Beverly Sackler School of Physics and Astronomy, Tel Aviv University, Tel Aviv 69978, Israel}

\begin{abstract}
\textcolor{black}{We investigate a two-dimensional tiling model. Even though
the degrees of freedom in this model are discrete, it has a hidden continuous global symmetry in the
infinite lattice limit, whose corresponding Goldstone modes are the quasicrystalline
phasonic degrees of freedom. 
We show that due to this continuous symmetry, and despite the apparent discrete nature of the model, a topological phase transition from a quasi-long-range ordered to a disordered phase occurs at a finite
temperature, driven by vortex proliferation.
We argue that some of the results are universal properties of two-dimensional systems whose ground state is a quasicrystalline state.}
\end{abstract}

\maketitle

\section{introduction}
\begin{figure*}[t]
\centering
%\float
(a){\includegraphics[width=0.26\textwidth]{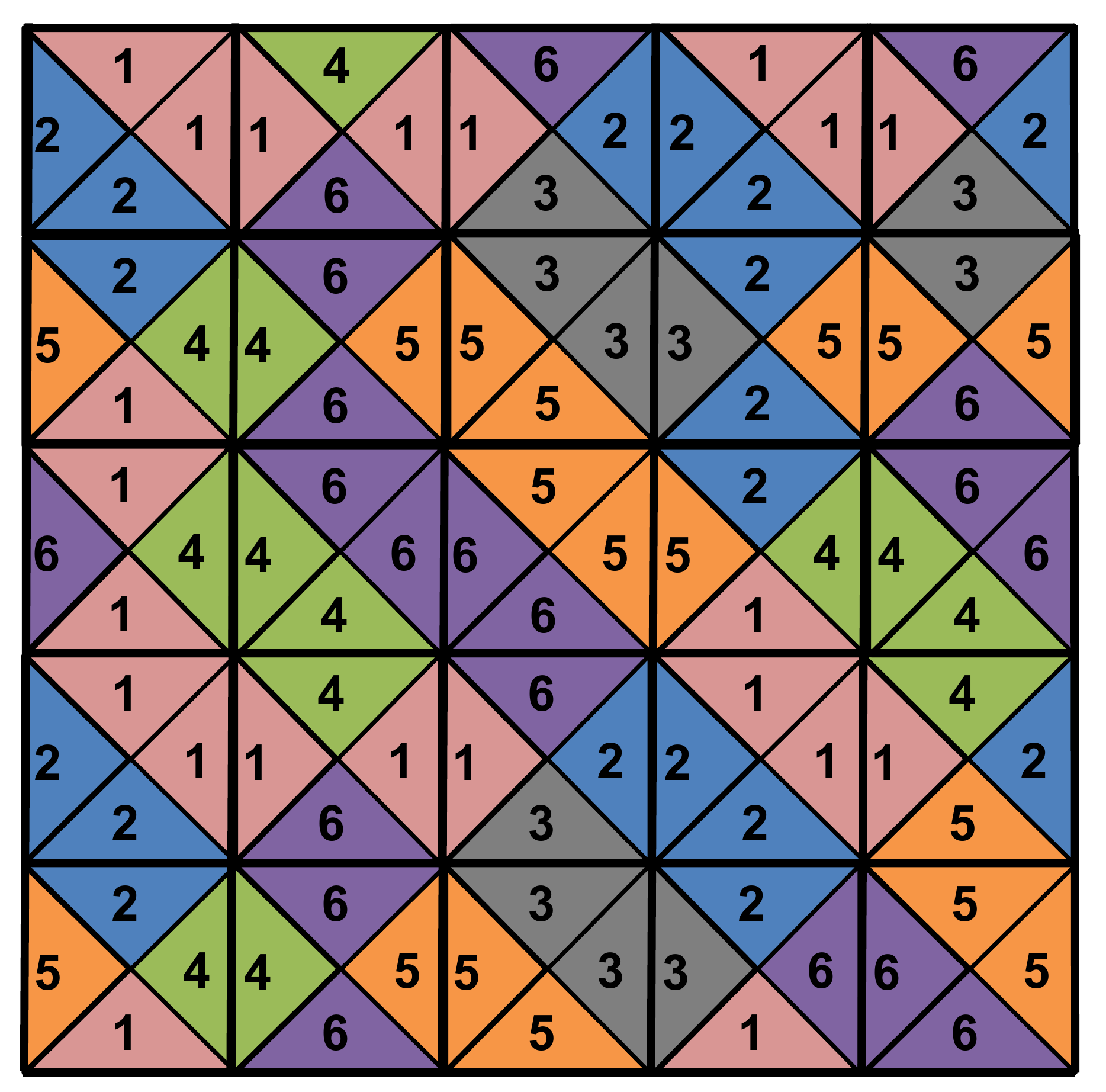}}
%\float
(b){
\includegraphics[width=0.26\textwidth]{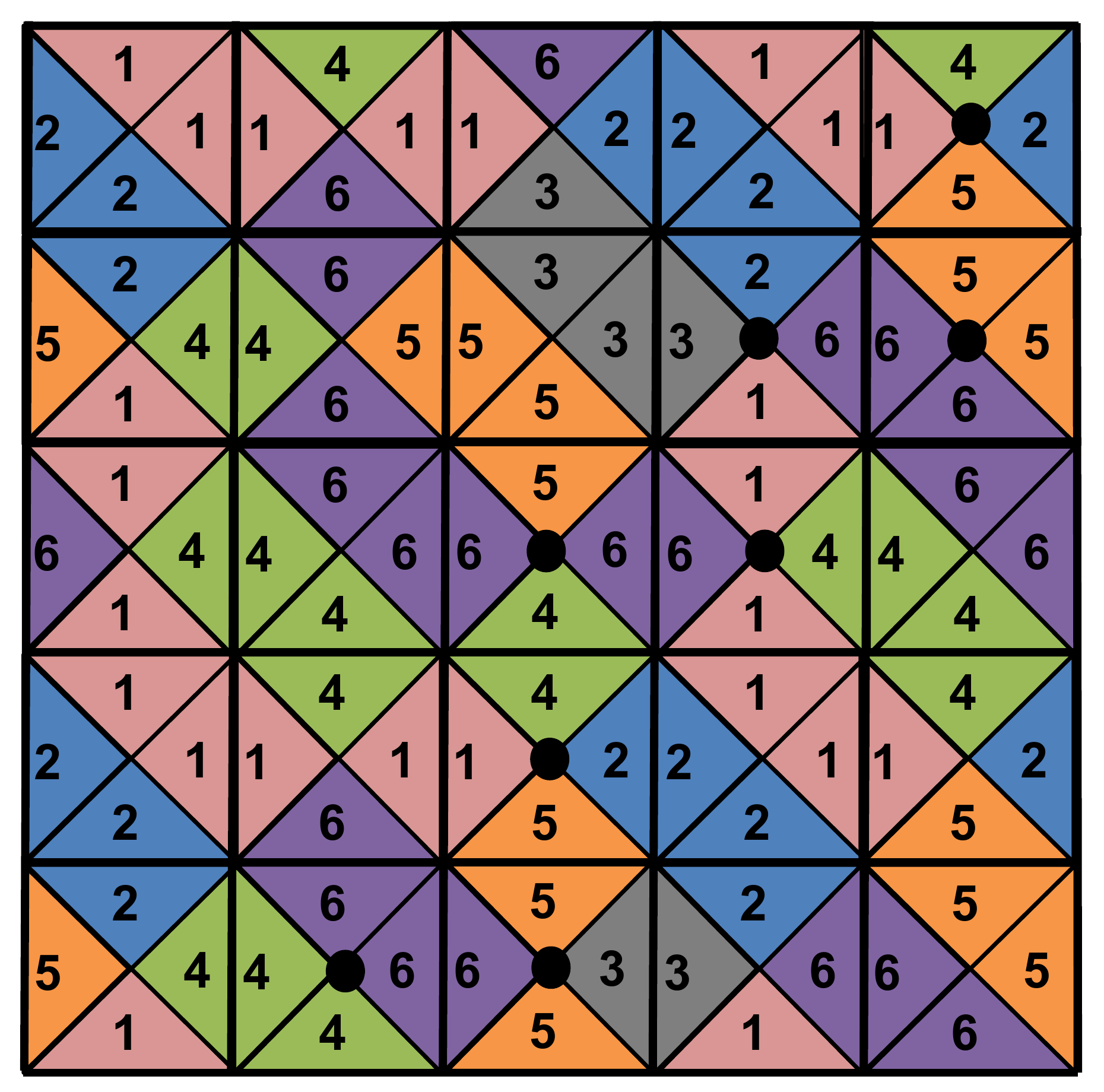}}
%\float(c)
{
\includegraphics[width=0.36\textwidth]{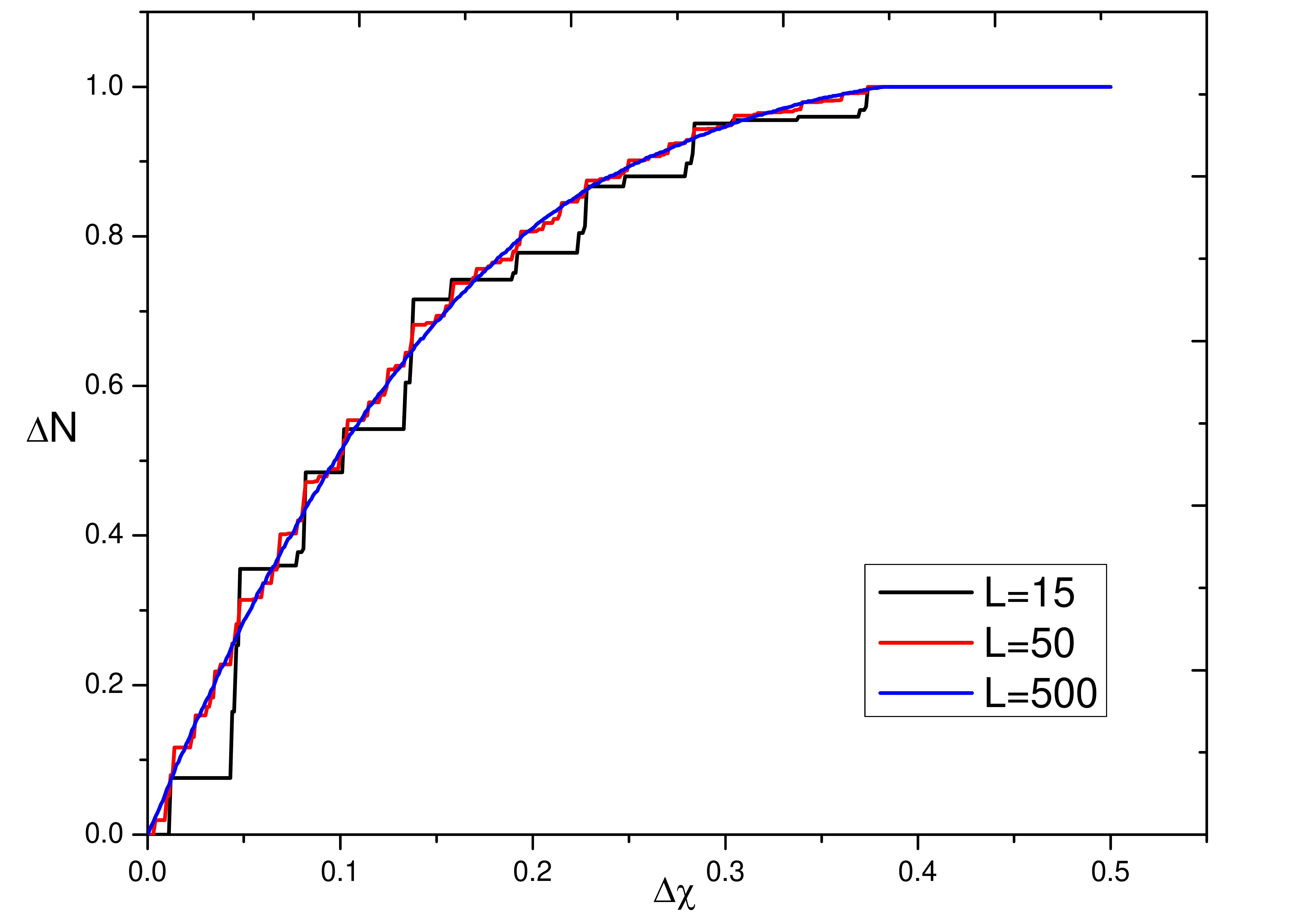}}
\caption{(Color online) (a) A perfect tiling configuration on a $5\times5$ lattice generate with $\chi_{1}=0.35$ and $\chi_{2}=0.6$. 
(b) The minimal change of $\chi_{2}$ required to change the $5\times5$ configuration shown in (a) is $\sim0.037$ leading to this tiling. The black dots at the center of the tiles denote tiles
that are changed in comparison with the configuration shown in (a). %(c) The configuration that corresponds
%to a maximal change of $\chi_{1}$ by $0.5$. All tiles are modified.
(c) $\Delta N$, the fraction of changed tiles as a function of the change $\Delta\chi$ of $\chi_{2}$, with respect
to the configuration $\chi_{1}=0.35,\chi_{2}=0.3$.}
\label{fig:domino}
\end{figure*}

\textcolor{black}{While the existence of quasicrystals \cite{Shechtman1984,Levine1984}
in nature is no longer debatable, it remains an open question if
materials can have a quasicrystalline ground state, and what the finite-temperature properties of this phase are \cite{DeBoissieu2012}. In addition, the characteristics of the phasonic degrees of freedom in quasicrystals are still the focus of much interest \cite{Widom2008, Kromer2012}. It is therefore
of great value to investigate the finite temperature physical properties of simple models with a quasicrystalline ground state. Such models can easily be constructed using the mathematical theory of tilings \cite{Grunbaum1986}, and have been extensively used for the study of quasicrystallinity \cite{Stadnik1999,Steinhardt1999,Suck2002,Trebin2006}. In particular, some finite temperature properties were studied using tiling models.
For example, the elastic properties of a three-dimensional model were shown to change upon a finite temperature phase transition \cite{Jeong1993,Dotera1994}, and a two-dimensional (2D) tiling model was recently shown to undergo a series of phase transitions leading from the quasicrystalline phase to the liquid phase through a number of intermediate periodic phases \cite{Nikola2013}.\\}
\begin{figure}[t]
\noindent \centering{}\textcolor{black}{\includegraphics[scale=0.3]{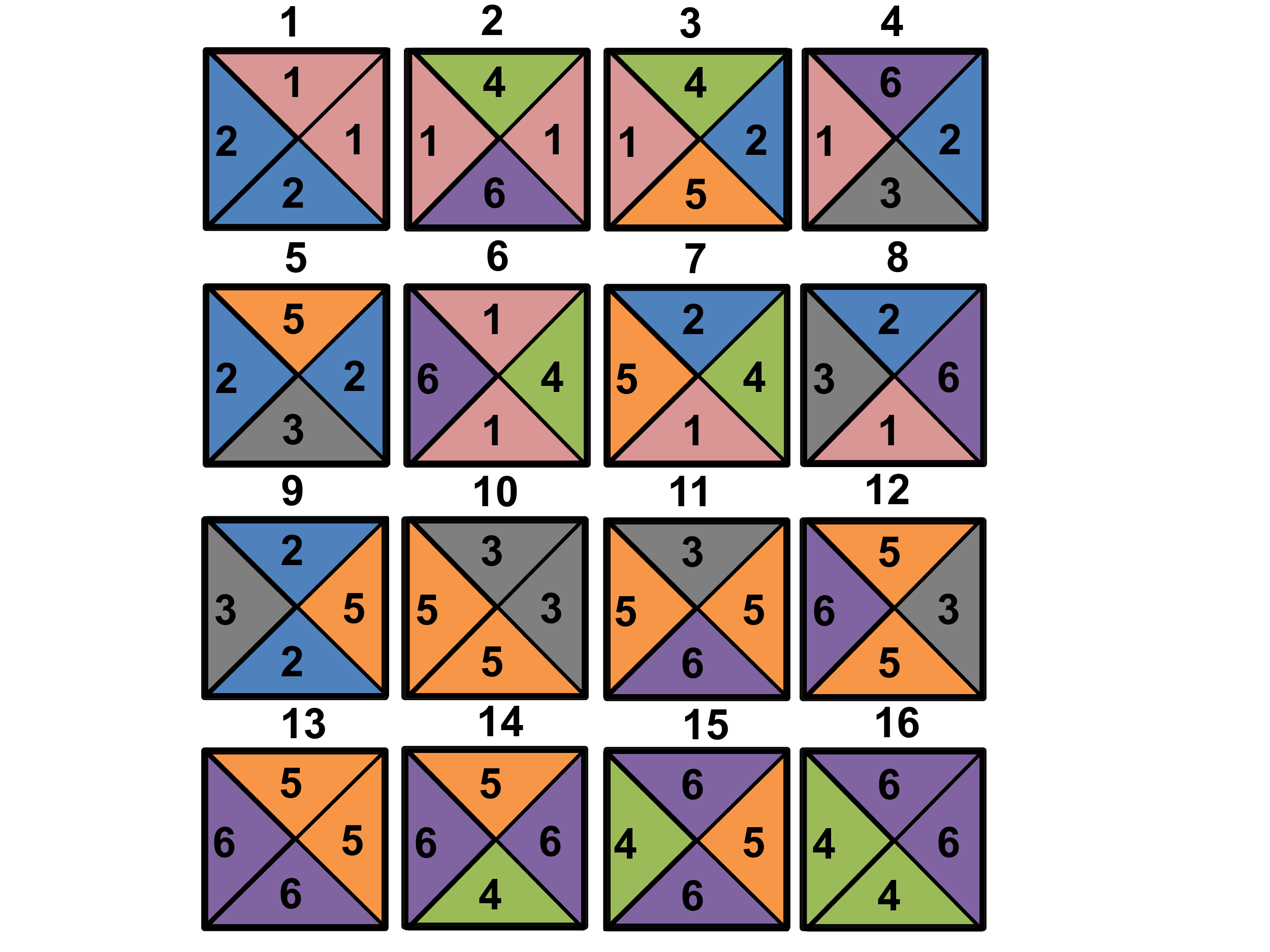}\caption{\label{fig:tiles} (Color online) The 16 Ammann tiles.}
}
\end{figure}
\textcolor{black}{The model studied here is based on the 16 Ammann
tiles, each of which being decorated with one label (out of possible six) on each of its four edges (Figure \ref{fig:tiles}).
Ammann \cite{Grunbaum1986} showed that
These tiles can perfectly tile the plane such that adjacent edges have matching labels. All such Domino-like tiling configurations are non-periodic and share a quasicrystalline order: well-defined Bragg peaks are observed in the Fourier transform of the densities of each given tile-type at frequencies incommensurate with the reciprocal lattice vectors. For an infinite system,
there is an uncountable number of different perfect tiling configurations, parameterized by two continuous phases, $\chi_{1},\chi_{2}\in [0,1)$ (see Appendix \ref{App:map}). These phases \cite{Mermin1992,Lifshitz2011}, are related to the amplitudes of the Bragg peaks (see equation (\ref{eq:fourier transform of tile density}) below). For any finite patch of a perfect tiling, these phases are not well defined, and can be described by fuzzy angles, whose
uncertainty is inversely proportional to the linear size $L$ \cite{Koch2009}. The number of different tilings of a finite system scales linearly with $N=L^2$. Accordingly, a finite change of $\chi_{1}$ and $\chi_{2}$ is required in order to induce any change in a finite patch tiling. However, a continuous change of $\chi_{1}$ and $\chi_{2}$ induces a continuous change in the {\it infinite} configuration: the fraction of tiles modified by an infinitesimal change of these phases is linear in this change (Figure \ref{fig:domino}). 
This hidden continuous symmetry of the perfect tilings is therefore manifestly non-local. }\
%\linebreak
In what follows, we show that this global continuous symmetry has a major
impact on the finite temperature behavior of the model studied
here. Namely, like a truly local continuous symmetry, it does not allow the system to be ordered at any positive temperature.

In order to study the model at a finite temperature, one needs to define the Hamiltonian. A natural choice, introduced by Leuzzi and Parisi \cite{Leuzzi2000}, is to identify the energy of a configuration with the number of mismatching edges. Thus, the (uncountably degenerate) ground states of the model are the perfect tilings exhibiting quasicrystalline order. We wish to study the stability of this order to thermal fluctuations. In order to  write the Hamiltonian in a convenient form, we define the 16-dimensional density vector $\overrightarrow{\rho}$, containing the 16 tile densities $\rho_{i}(\mathbf{r})$,
each of which is a unity if the tile at $\mathbf{r}$ is of type $i$, and zero otherwise. In terms of these, the Hamiltonian takes the form 
\begin{equation}
H=\sum_{\mathbf{r}}\left[\overrightarrow{\rho}{}^{\dagger}(\mathbf{r})Y\overrightarrow{\rho}(\mathbf{r}+\mathbf{\hat{y}})+\overrightarrow{\rho}{}^{\dagger}(\mathbf{r})X\overrightarrow{\rho}(\mathbf{r}+\hat{\mathbf{x}})\right],\label{eq:energy}
\end{equation}
where $X$ and $Y$ are known interaction matrices, dictated by the above edge-matching rule, whose explicit form can be found in Appendix \ref{App:XY}. The unit vectors $\mathbf{\hat{x}}$ and $\mathbf{\hat{y}}$ connect each site to (two of) its nearest neighbors. Note that the lattice constant is chosen as the length unit.

Note that in a general quasicrystalline system two kinds of gapless collective excitations exist: phonons and phasons
\cite{Socolar1986, Lifshitz2011}. Phonons describe locally uniform translations, while phasons describe correlated rearrangements of atoms. Our model is defined on a fixed lattice and therefore the low-energy excitations described by this model are the phasonic degrees of freedom.

\textcolor{black}{Previous works \cite{Leuzzi2000,Koch2009,Rotman2011}
have provided numerical evidence that the model undergoes a symmetry
breaking phase transition from a quasicrystalline low temperature
phase into a high temperature disordered phase. This seems to contradict
the well known Mermin-Wagner theorem, stating that continuous symmetries cannot be spontaneously broken in 2D (or one-dimensional)
systems \cite{Mermin1966,Hohenberg1967,Mermin1968}. However, the theorem relies on the existence of a {\it local}
order parameter field that can be changed continuously. In our case, each tile, and even the ground state of each finite patch, has a finite degeneracy and cannot be changed continuously. A slow gradient of $\chi_{1}$ and $\chi_{2}$ will not make any change in most finite patches of the system, and will be manifested by a discrete jump in the energy for some isolated patches. It is therefore not clear whether the Mermin-Wagner theorem applies here.}
\section{Absence of quasicrystalline order}
\textcolor{black}{We first provide an argument that quasicrystallinity is broken at any finite temperature, in a fashion similar to the case of a truly continuous local symmetry. For this purpose, we assume that the system is ordered at low temperatures, and self consistently calculate its finite temperature properties. It is then shown that thermal excitations destroy the order.}\

\textcolor{black}{The global symmetry is reflected in the
Fourier transform of the tile densities. %We define the 16-dimensional density vector $\overrightarrow{\rho}$, containing the 16 tile densities $\rho_{i}(\mathbf{r})$,
%each of which is a unity if the tile at $\mathbf{r}$ is of type $i$, and zero otherwise. 
At a ground state characterized by the two phases $\chi_1,\chi_2$, the Fourier transform of $\overrightarrow{\rho}(\mathbf{r})$, $\overrightarrow{\psi}(\mathbf{q})$, takes the form 
%\begin{equation}
%\overrightarrow{\psi}(\mathbf{q})=
%N\sum_{m,n,i,j}\delta(\mathbf{q}-2\pi(n\tau\mathbf{\hat{x}}+m\tau\mathbf{\hat{y}}+i\mathbf{\hat{x}}+j\mathbf{\hat{y}}))e^{2\pi i\left(n\chi_{1}+m\chi_{2}\right)}\overrightarrow{\psi_{0}}(n,m),\label{eq:fourier transform of tile density}
%\end{equation}
\begin{eqnarray}
\overrightarrow{\psi}(\mathbf{q}) & = & N\sum_{m,n,i,j}\delta(\mathbf{q}-2\pi(n\tau\mathbf{\hat{x}}+m\tau\mathbf{\hat{y}}+i\mathbf{\hat{x}}+j\mathbf{\hat{y}}))\nonumber\\
& & \times e^{2\pi i\left(n\chi_{1}+m\chi_{2}\right)}\overrightarrow{\psi_{0}}(n,m), \label{eq:fourier transform of tile density}
\end{eqnarray}
where $\chi_{1}$ and $\chi_{2}$ are
the continuous phases discussed above, $\tau=\frac{\sqrt{5}-1}{2}$ is the
inverse golden ratio, and $\overrightarrow{\psi_{0}}(n,m)$ are analytically calculated constant
amplitudes (see Appendix \ref{App:fourier components}). Note that $\mathbf{q}$ is defined modulo reciprocal lattice vectors
 $G=2\pi(i\mathbf{\hat{x}}+j\mathbf{\hat{y}})$. Bragg peaks are thus spanned by 4 independent basis reciprocal vectors (like the closely related square Fibonacci tiling \cite{Lifshitz2002}), consistent with the quasicrystalline nature of the model.}

Assuming low temperature quasicrystalline order, only long wavelength
excitations should be considered. At scales smaller
than the typical wavelength of the contributing excitations, the system
appears ordered, slowly passing from one local ground
state to another. To express this idea formally, we define the local
Fourier transform of a function $f(\mathbf{x})$ as 
\begin{equation}
f(\mathbf{x},\mathbf{k})=\frac{1}{A}\sum_{\mathbf{x'}}f(\mathbf{x'})e^{-i\mathbf{k}\cdot\mathbf{x}'}w_{\sigma}(\mathbf{x-x'}),\label{eq:local fourier transform}
\end{equation}
where $w_{\sigma}$ is a weight function with a finite length scale
$\sigma$ and $A=\sum_{\mathbf{x}}w_{\sigma}(\mathbf{x})$. This weight
function makes sure that we take only contributions around the point
$\mathbf{x}$. In order to simplify the analysis we take $w_{\sigma}$ to be
unity in some region with a length scale $\sigma\gg1$ around the origin,
and zero otherwise. As long as $\sigma$ is large
enough, the shape of this region is not important. 

We now consider long wavelength excitations where $\chi_{1}(\mathbf{r})$ and $\chi_{2}(\mathbf{r})$ change slowly with $\mathbf{r}$, being approximately constant on length scale $\sigma$. As the system appears locally ordered, its local Fourier transform is 
%\begin{equation}
%\overrightarrow{\psi}(\mathbf{q})=
%N\sum_{m,n,i,j}\delta(\mathbf{q}-2\pi(n\tau\mathbf{\hat{x}}+m\tau\mathbf{\hat{y}}+i\mathbf{\hat{x}}+j\mathbf{\hat{y}}))e^{2\pi i\left(n\chi_{1}(\mathbf{r})+m\chi_{2}(\mathbf{r})\right)}\overrightarrow{\psi_{0}}(n,m),\label{eq:local fourier transform of small excitation}
%\end{equation}
\begin{eqnarray}
\overrightarrow{\psi}(\mathbf{q}) & = & N\sum_{m,n,i,j}\delta(\mathbf{q}-2\pi(n\tau\mathbf{\hat{x}}+m\tau\mathbf{\hat{y}}+i\mathbf{\hat{x}}+j\mathbf{\hat{y}}))\nonumber\\
 & & \times e^{2\pi i\left(n\chi_{1}(\mathbf{r})+m\chi_{2}(\mathbf{r})\right)}\overrightarrow{\psi_{0}}(n,m),
\label{eq:local fourier transform of small excitation}
\end{eqnarray}
where $\chi_{1}(\mathbf{r})$ and $\chi_{2}(\mathbf{r})$ are the
phases corresponding to the local ground state. 
%We rewrite the Hamiltonian in a matrix form 
%\begin{equation}
%H=\sum_{\mathbf{r}}\left[\overrightarrow{\rho}{}^{\dagger}(\mathbf{r})Y\overrightarrow{\rho}(\mathbf{r}+a\mathbf{\hat{y}})+\overrightarrow{\rho}{}^{\dagger}(\mathbf{r})X\overrightarrow{\rho}(\mathbf{r}+a\hat{\mathbf{x}})\right],\label{eq:energy}
%\end{equation}
%where $X$ and $Y$ are known interaction matrices. 
In terms of the local Fourier transform, the 
Hamiltonian (Eq. \ref{eq:energy}) then takes the form
\begin{eqnarray}
H & = & \sum_{\mathbf{r},\mathbf{k}}\left[\overrightarrow{\psi}{}^{\dagger}(\mathbf{r},\mathbf{k})Y\overrightarrow{\psi}(\mathbf{r}+\mathbf{\hat{y}},\mathbf{k})e^{ik_{y}}+\right.\nonumber \\
 &  & \left.\overrightarrow{\psi}{}^{\dagger}(\mathbf{r},\mathbf{k})X\overrightarrow{\psi}(\mathbf{r}+\hat{\mathbf{x}},\mathbf{k})e^{ik_{x}}\right].
\end{eqnarray}
For a locally ordered configuration, one can plug in the local ground state approximation, equation
(\ref{eq:local fourier transform of small excitation}), and get the effective long wavelength Hamiltonian, 
$E[\chi_{1}(\mathbf{r}),\chi_{2}(\mathbf{r})]$
\begin{eqnarray}
E & = & \sum_{\mathbf{r}}\sum_{m,n}\left(e^{2\pi i\left(n\partial_{x}\chi_{1}+m\partial_{x}\chi_{2}\right)}\overrightarrow{\psi_{0}^{\dagger}}(n,m)X\overrightarrow{\psi_{0}}(n,m)e^{i2\pi n\tau}+\right.\nonumber \\
 &  & \left.\text{ }e^{2\pi i\left(n\partial_{y}\chi_{1}+m\partial_{y}\chi_{2}\right)}\overrightarrow{\psi_{0}^{\dagger}}(n,m)Y\overrightarrow{\psi_{0}}(n,m)e^{i2\pi m\tau}\right),
\end{eqnarray}
where $\partial_{x}\chi_i=\chi_i(\mathbf{r}+\hat{\mathbf{x}})-\chi_i(\mathbf{r})$
and $\partial_{y}\chi_i=\chi_i(\mathbf{r}+\hat{\mathbf{y}})-\chi_i(\mathbf{r})$
are the discrete derivatives of $\chi_i$. The sums over $m$ and $n$ can be performed
numerically, and the final result, to lowest order in the derivatives,
is 
\begin{eqnarray}
E & =\sum_{\mathbf{r}} & A\left(\left|\partial_{x}\chi_{1}\right|+\left|\partial_{y}\chi_{2}\right|\right)+B\left(\left|\partial_{x}\chi_{2}\right|+\left|\partial_{y}\chi_{1}\right|\right)\label{effective field theory}\\
 &  & +C\left(\left|\partial_{x}\chi_{1}+\tau\partial_{x}\chi_{2}\right|+\left|\partial_{y}\chi_{2}+\tau\partial_{y}\chi_{1}\right|\right)\nonumber \\
 &  & +D\left(\left|\partial_{y}\chi_{1}+\tau\partial_{y}\chi_{2}\right|+\left|\partial_{x}\chi_{2}+\tau\partial_{x}\chi_{1}\right|\right),\nonumber 
\end{eqnarray}
where $A\approx1.00,B\approx1.94,C\approx1.57,D\approx0.61$. 

We now investigate this effective Hamiltonian at finite temperatures.
Once we rephrased the low-T physics of the model in terms of truly continuous fields,
it is rather obvious that the Mermin-Wagner theorem applies, and thermal excitations
must destroy the order in any finite temperature. However, three notes are in order. First, note that the Mermin Wagner theorem holds even though the effective field theory is non-analytic, as long as it is continuous \cite{Ioffe2002}. Second, the transformation from the tiles degrees of freedom to the continuous phases involves a non trivial, singular, Jacobian, which at finite temperature translates into a complicated entropic term. While this entropic term remains unspecified, it must preserve the continuous symmetry in the local ground state approximation, and therefore should not affect our argument. Third, as discussed above, the local phases are never truly continuous. Each finite patch of the system has a finite ground state degeneracy, and thus the number of distinct values that any of the phases can take is finite and scales like the patch size $\sigma$, which can be taken to be the largest scale over which the system is in an approximate ground state. However, here one can invoke the discrete tile picture of the system: as any mismatch in a tiling costs at least one unit of energy, the density of mismatches at low temperatures is at most $O(\exp(-\Delta/T))$ with $\Delta$ of order unity. Thus, $\sigma$, the scale upon which the system is at a local ground state, can be made exponentially large as temperature drops down, and the system can
be effectively described by continuous phases.
The assumption of low temperature order leads to a contradiction, and the system is therefore not ordered at any finite temperature.

The transition found in \cite{Leuzzi2000,Koch2009,Rotman2011}
is therefore not a symmetry breaking transition. We now turn to investigate its
true nature. 
\begin{figure}[t]
\noindent \centering{} \textcolor{black}{\includegraphics[scale=0.31]{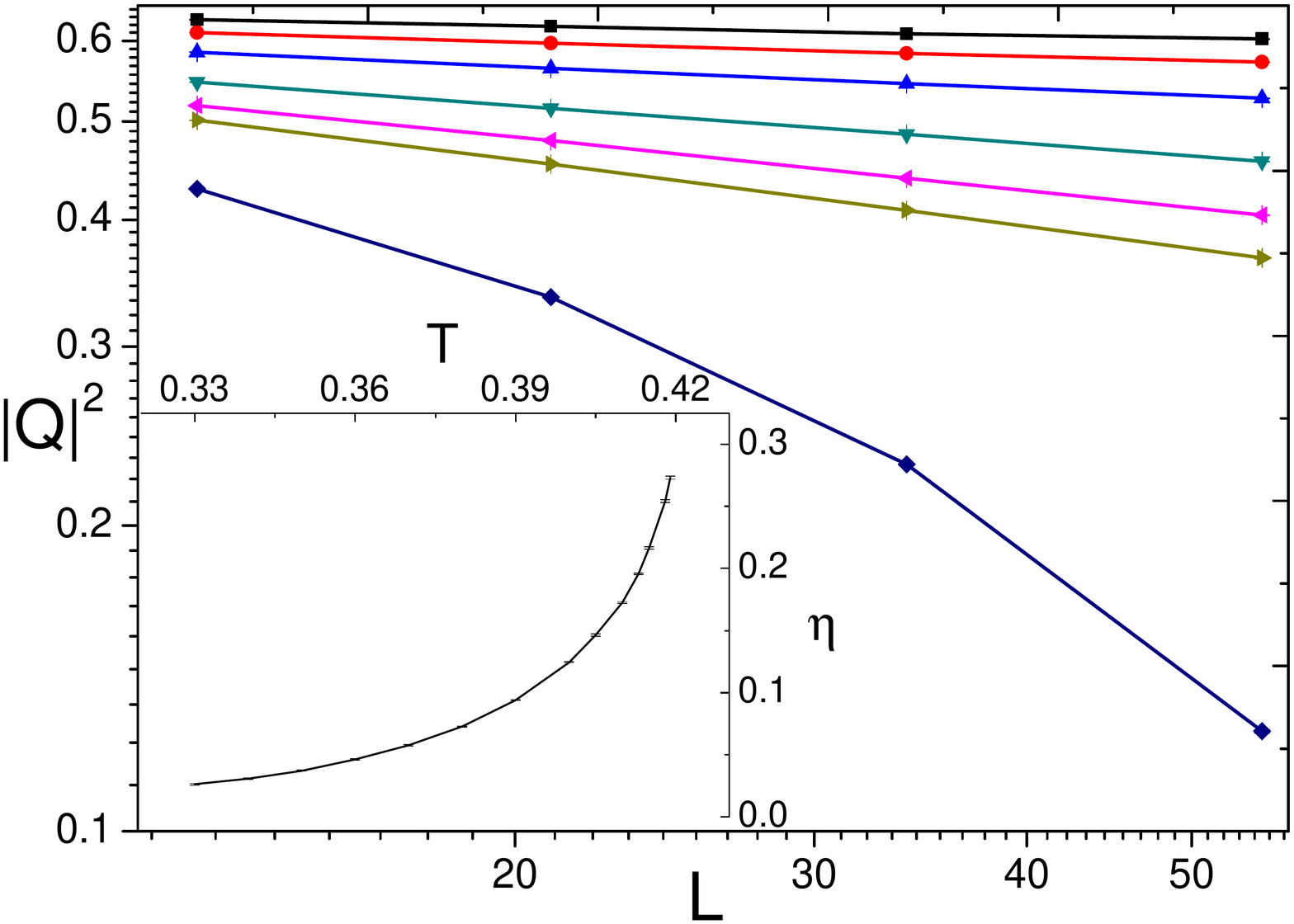}
\caption{\label{fig:tiling_m2} (Color online) $\left|Q\right|^{2}=\frac{1}{2}\left(\left|Q_{x}\right|^{2}+\left|Q_{y}\right|^{2}\right)$
 as a function of the system's size different temperatures (top to bottom: $T=0.34,0.36,0.38,0.4,0.41,0.415,0.43$). Note the log-log scale. 
For $T$ very close but higher than $T_{c}\sim0.42$, an exponential decay
is evident. Inset: $\eta$ as a function of $T$ below the critical
temperature. }
}
\end{figure}
\section{Finite temperature behavior of the model}
\textcolor{black}{
 A natural choice for the order parameters of the model, closely
related to the one defined in \cite{Rotman2011}, is the Fourier coefficients of the tile densities at
the basis reciprocal vectors: 
\begin{equation}
q_{i}^{x}=\frac{1}{N}\sum_{\mathbf{r}}e^{-i2\pi\tau x}\rho_{i}(\mathbf{r})\text{, }q_{i}^{y}=\frac{1}{N}\sum_{\mathbf{r}}e^{-i2\pi\tau y}\rho_{i}(\mathbf{r}),\label{eq:naive order parameter-1}
\end{equation}
where $i$ is one of the 16 tile-types, and its choice is arbitrary. Note the need for
two order parameters, as the ground state manifold is parameterized
by two phases. While this form is correct, a more symmetric
and numerically preferable generalization is
\begin{equation}
Q_{x}=\frac{1}{N}\sum_{i,\mathbf{r}}e^{-i2\pi\tau x}e^{i\gamma_{i}^{x}}\rho_{i}(\mathbf{r})\text{,  }Q_{y}=\frac{1}{N}\sum_{i,\mathbf{r}}e^{-i2\pi\tau y}e^{i\gamma_{i}^{y}}\rho_{i}(\mathbf{r}),\label{eq:order parameter-1}
\end{equation}
which sums the contribution from all tile-types $i$. The phases $\gamma_{i}^{x}$ and $\gamma_{i}^{y}$ are the relative phases between the Bragg peaks amplitudes observed for each tile-type (see Appendix \ref{App:fourier components}).
We measured these order parameters in the vicinity of the transition ($T_c\simeq0.418$) and below it, using  Monte-Carlo simulations of the {\it{original tiling model}}.
Ground state configurations are nearly periodic with periodicities that are Fibonacci numbers (see Appendix \ref{App:map}).
We found that finite size effects are minimized using periodic boundary conditions, provided linear system size is a Fibonacci number.}

\textcolor{black}{
Well below the transition $\left|Q\right|^{2}\propto L^{-\eta(T)}$ (Figure \ref{fig:tiling_m2}), implying a power law decay of the correlation function with the
same exponent $\eta(T)$. 
Above the transition $\left|Q\right|^{2}$ falls
exponentially, indicating short-range correlations. 
This resembles the situation in the XY model, for example, which exhibits a quasi-long-range order (QLRO) at low-T and a topological Kosterlitz-Thouless (KT) transition \cite{Kosterlitz1972,Kosterlitz1974} to a disordered phase.}

The KT transition is associated with vortex
unbinding. It is therefore natural to ask whether single vortices
become stable at high temperatures in our system. A vortex in the
field $\chi_{1}$, for example, is given by a configuration associated with 
\begin{equation}
\chi_{1}=\frac{1}{2\pi}\arctan\frac{y}{x},\chi_{2}=0.\label{eq:vortex solution}
\end{equation}

\begin{figure}[t]
\noindent \centering{}\textcolor{black}{\includegraphics[scale=0.45]{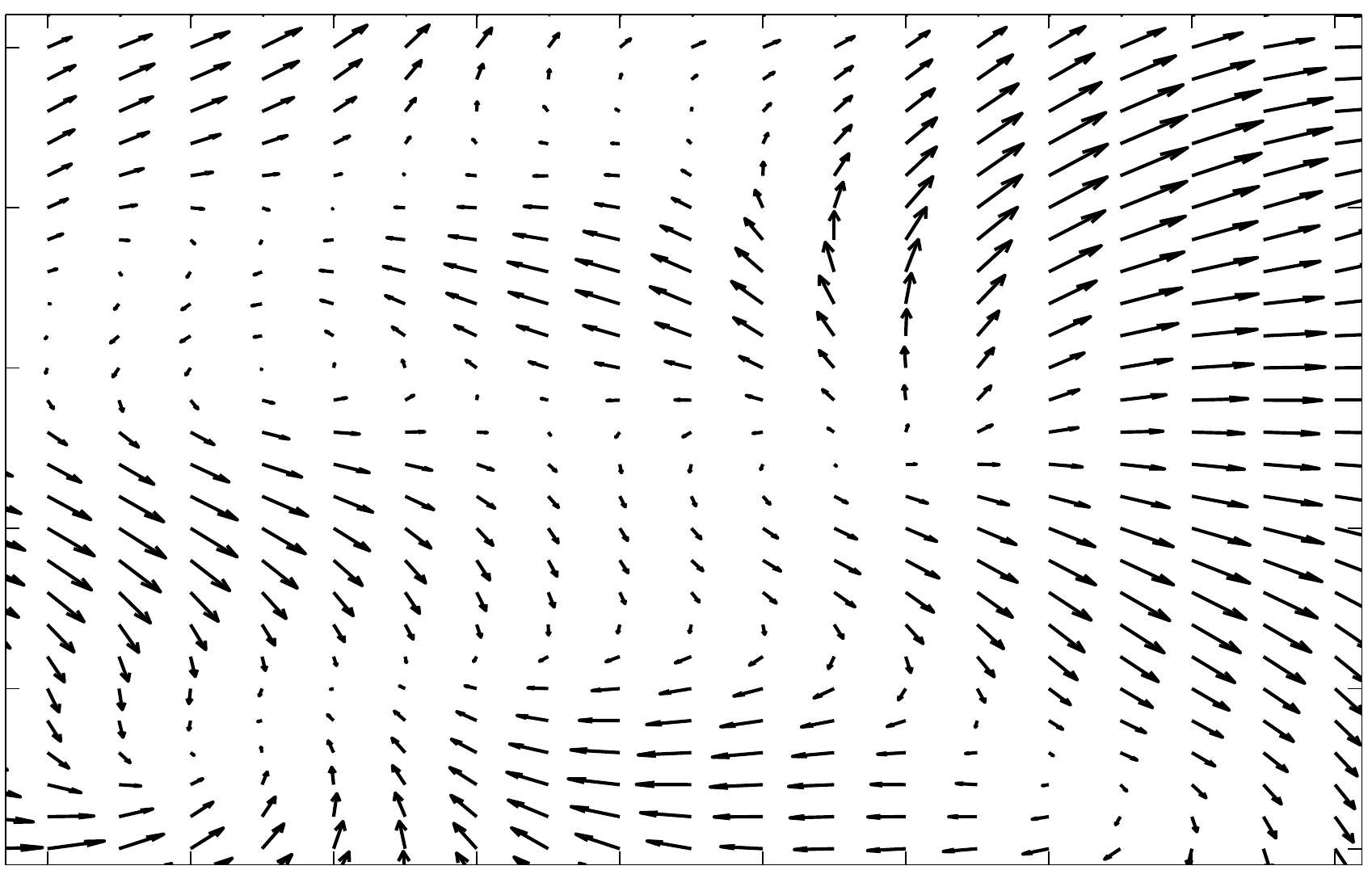}\caption{\label{fig:vortex} A typical configuration above the transition. The arrows represent the complex numbers $\psi_{1}(\mathbf{r},2\pi\tau\mathbf{\hat{x}})$ with $w_{\sigma}=\exp(-r^2/\sigma^2)$ and $\sigma=5$. We note that vortices similar to those shown in the figure were observed in all the configurations above $T_c$. Below $T_c$, no vortices were observed.}
}
\end{figure}

Using equation (\ref{effective field theory}), it is easy to see that the energy of the vortex diverges, $H_{vortex}\propto L$. The positional entropy of the vortex, on the other hand, grows with system's size only as $\log L$. A naive application of the standard KT argument, explaining the onset of vortex unbinding as a consequence of the positional entropy
overcoming the energy, would lead to the conclusion that in our case energy always
wins and vortices are
never stable. This conclusion is clearly wrong - our transition is associated with proliferation of vortices, 
see figure \ref{fig:vortex}. Upon integrating-out the fast degrees of freedom, the field theory (\ref{effective field theory}) is likely to be renormalized into an effective gaussian free-energy functional, which results in a logarithmically-diverging vortex effective energy. This was shown to be the case for a similar tiling model at any finite temperature \cite{Tang1990}. %It is therefore likely that upon integrating-out the fast degrees of freedom, the field theory (\ref{effective field theory}) is renormalized into
%an effective gaussian free-energy functional, which results in a logarithmic
%vortex effective energy. 
The standard KT argument for positional entropy overcoming the {\it effective} free-energy of the vortex at high temperatures does hold, and vortex unbinding will drive a topological phase transition. \\
\textcolor{black}{In the usual KT scenario, where
the energy itself is quadratic to lowest order, $\eta\propto T$ at
low temperatures. In contrast, in our case $\eta$
shows a highly non-linear behavior (Figure \ref{fig:tiling_m2}, inset).
This too signals that the effective coupling constant is strongly renormalized, and becomes temperature dependent. 
The very steep decay of $\eta$ as one moves away from
the transition implies that in finite lattices at low temperatures,
the system appears to be ordered, and the identification of the algebraic
correlations is very difficult in reasonably sized systems. This explains why previous works \cite{Leuzzi2000,Koch2009,Rotman2011} identified the low temperature phase as an ordered one.\
}

However, one feature of our transition deviates from the KT scenario. At the KT transition, the heat capacity $C_v$ exhibits a weak (numerically undetectable) $C_\infty$ essential singularity. As shown in \cite{Leuzzi2000,Koch2009,Rotman2011}, the tiling model exhibits a distinctive sharp peak in $C_v$ at $T_c$. In particular, we observed (for $L$ up to $89$) a clear power-law divergence $\left.\frac{dC_{V}}{dT}\right|_{T_{C}}\propto L^{-\epsilon}$ with $\epsilon$ (very roughly) $\sim 0.5(3)$, indicating a finite-order transition with $\nu\simeq1.25(20)$. Bearing in mind the large uncertainties in these numerical estimates, and the limited system sizes, this seems to suggest our transition may be of a different universality class than the standard KT transition. A qualitative change in critical behavior due to interaction between two XY fields was pointed out in the context of a double-layer XY model \cite{Parga1980}.
\\
It is worth saying a few words about the form of
the long wavelength Hamiltonian, equation (\ref{effective field theory}).
Usually, only analytic terms are considered when one constructs an effective field theory. The tiling model provides
an example where non-analytic terms arise naturally from first principles. In fact, it was already suggested that terms of the form $\sim\left|\partial\chi\right|$ describe the energy of phasons in general systems with a quasicrystalline
ground state \cite{Socolar1986}. A phase in which the {\it{free energy}} is characterized by such a non-analytic form is usually referred to as a locked phase \cite{Steinhardt1999}. In 3D, one expects to find a finite temperature transition from this phase to an unlocked phase, characterized by a quadratic free energy \cite{Jeong1993,Dotera1994}. However, in 2D systems, such as the one studied in this work, the transition occurs at zero temperature, as was shown in \cite{Tang1990}. A similar derivation of the energy can be made for an
analog three-dimensional system, where we expect that the equivalent of (\ref{effective field theory}) would be the
relevant low-temperature effective theory.\\ 
%As we have seen, at positive temperatures these non-analytic terms are heavily renormalized by thermal fluctuations, and flow to the analytic quadratic free energy.
%While this is true in our 2D case, similar derivation can be made for an
%analog three-dimensional system, where it is possible that the equivalent of (\ref{effective field theory}) would be the
%relevant low-temperature effective theory \cite{Jeong1993,Dotera1994}.\\}
\section{Conclusions}
\textcolor{black}{We described here a topological phase transition in a system with discrete degrees of freedom. It is constructive to juxtapose this behavior with a similar scenario. 
The clock model, where each spin can take one of $q$ possible planar directions \cite{Jose1977, Elitzur1979,Ortiz2012}, exhibits a KT transition for $q>4$. In this case, as long as $k_BT$ exceeds the energy of rotating a single spin between two neighboring directions, thermal fluctuations restore the continuous U(1) symmetry and one effectively gets back an XY model with 
algebraically decaying correlations. Indeed, as temperature lowers the discrete nature is revealed, and a second phase transition occurs below which the system is ordered.
In comparison, in our tiling model the hidden continuous U(1) symmetry is restored not by temperature but rather by going into larger and larger finite ordered patches. The lower the temperature, the larger are the ordered patches in the system, and thus the QLRO phase survives for arbitrarily low temperatures. Given that the continuous symmetry discussed here is a general property of quasicrystals, similar arguments may lead to the conclusion that any 2D model with a quasicrystalline ground state (with either discrete or continuous degrees of freedom), cannot be ordered at any positive temperature. Indeed, algebraic correlations (but not a KT transition) were observed in a Penrose tiling model \cite{Tang1990} and various random tiling models \cite{Steinhardt1999}, and we expect the tiling model recently studied by Nikola {\it et al} \cite{Nikola2013} to exhibit QLRO at low temperatures as well. \textcolor{black}{ Furthermore, in our model rotational symmetry is explicitly broken by the underlying real-space lattice. However, the above formulation of the configuration in terms of the local phases enables one to study, in off-lattice models, the orientational QLRO of these fields. One expects a two-step melting of the QLRO quasicrystal through an intermediate "Hexatic" (or, rather, "Pentatic" for a five-fold symmetric quasicrystal) phase, as was predicted in \cite{De1989}.}}
\begin{acknowledgments}
We would like to acknowledge Ron Lifshitz, Giorgio Parisi, and Moshe Schwartz for many helpful discussions, as well as Ziv Rotman for helping us with the numerical simulations.
\end{acknowledgments}
\appendix

\section{Generating Ground State Configurations and Excitations \label{App:map}}

\begin{figure}[t]
\includegraphics[scale=0.45]{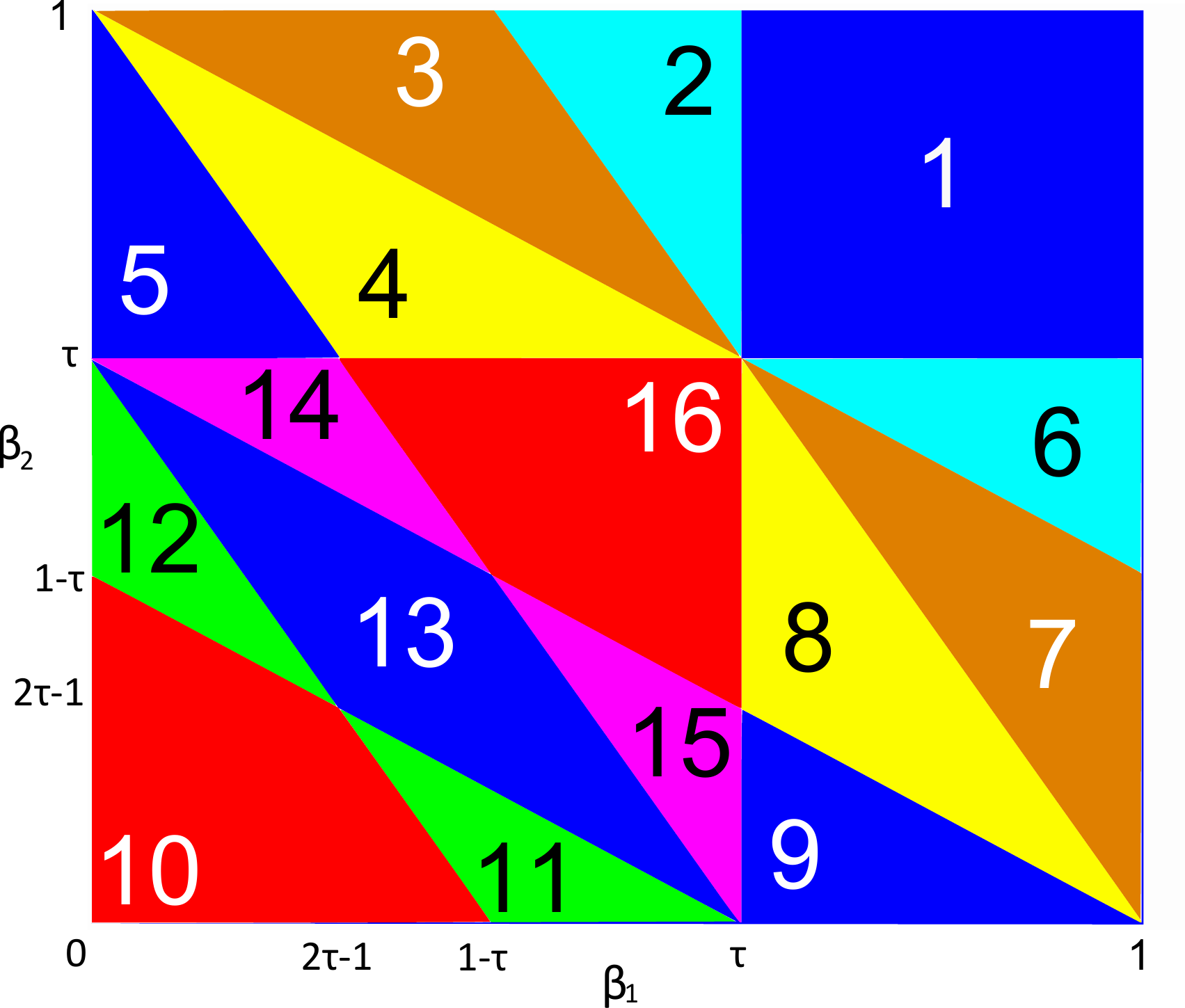}\label{map}\caption{(Color online) The mapping between the $\beta_{1}-\beta_{2}$ torus and the 16 tile types.}
\end{figure}

For many purposes (some of which will be mentioned soon), it is necessary
to generate a configuration with given $\chi_{1}$ and $\chi_{2}$. In what follows, we show how this can
be done. Motivated by the connection between our model and the square
Fibonacci sequence \cite{Koch2009}, we define the two functions 
\begin{equation}
\beta_{1}(\mathbf{r})=\left\{ \chi_{1}+\tau x\right\} ,\beta_{2}(\mathbf{r})=\left\{ \chi_{2}+\tau y\right\} ,\label{eq:beta}
\end{equation}
where $\left\{\cdots\right\}$ is the fractional part of $\cdots$, i.e. $(\cdots)mod1$. In a perfect tiling, each tile
type is associated with a given region on the $\beta_{1}-\beta_{2}$ torus (figure (\ref{map})).
Together with equation (\ref{eq:beta}), this mapping allows for generating a perfect tiling for an arbitrary choice of $\chi_{1}$
and $\chi_{2}.$
In order to study the excitations within the local ground state approximation,
one may now construct non-perfect configurations with slowly varying
phases: $\beta_{1}(\mathbf{r})=\left\{ \chi_{1}(\mathbf{r})+\tau x\right\} ,\beta_{2}(\mathbf{r})=\left\{ \chi_{2}(\mathbf{r})+\tau y\right\} $.
Looking at the associated non-perfect tiling configuration, it is
possible to verify numerically that their energy follows the relation 
\begin{eqnarray}
E & =\sum_{\mathbf{r}} & A\left(\left|\partial_{x}\chi_{1}\right|+\left|\partial_{y}\chi_{2}\right|\right)+B\left(\left|\partial_{x}\chi_{2}\right|+\left|\partial_{y}\chi_{1}\right|\right)\label{eq:H}\\
 &  & +C\left(\left|\partial_{x}\chi_{1}+\tau\partial_{x}\chi_{2}\right|+\left|\partial_{y}\chi_{2}+\tau\partial_{y}\chi_{1}\right|\right)\nonumber \\
 &  & +D\left(\left|\partial_{y}\chi_{1}+\tau\partial_{y}\chi_{2}\right|+\left|\partial_{x}\chi_{2}+\tau\partial_{x}\chi_{1}\right|\right),\nonumber 
\end{eqnarray}
which was derived in the main text (to lowest order in derivatives). 

Using this explicit tiling construction, it is easy to see the near periodicities of the perfect tiling configurations.
Fibonacci numbers satisfy $\tau F_{n}\simeq F_{n-1}$
(for large $n$), and thus $\beta_{1}(x,y)\backsimeq\beta_{1}(x+F_{n},y)$
and $\beta_{2}(x,y)\backsimeq\beta_{2}(x,y+F_{n})$.
\section{Explicit form of X and Y \label{App:XY}}
The matrices X and Y used to define the Hamiltonian (\ref{eq:energy}) are derived directly from the matching rules and the definition of the $16$ tiles found in figure \ref{fig:tiles}. Their explicit form is: 
$$Y=\left(
\begin{array}{cccccccccccccccc}
 1 & 1 & 1 & 1 & 1 & 1 & 0 & 0 & 0 & 1 & 1 & 1 & 1 & 1 & 1 & 1 \\
 1 & 1 & 1 & 0 & 1 & 1 & 1 & 1 & 1 & 1 & 1 & 1 & 1 & 1 & 0 & 0 \\
 1 & 1 & 1 & 1 & 0 & 1 & 1 & 1 & 1 & 1 & 1 & 0 & 0 & 0 & 1 & 1 \\
 1 & 1 & 1 & 1 & 1 & 1 & 1 & 1 & 1 & 0 & 0 & 1 & 1 & 1 & 1 & 1 \\
 1 & 1 & 1 & 1 & 1 & 1 & 1 & 1 & 1 & 0 & 0 & 1 & 1 & 1 & 1 & 1 \\
 0 & 1 & 1 & 1 & 1 & 0 & 1 & 1 & 1 & 1 & 1 & 1 & 1 & 1 & 1 & 1 \\
 0 & 1 & 1 & 1 & 1 & 0 & 1 & 1 & 1 & 1 & 1 & 1 & 1 & 1 & 1 & 1 \\
 0 & 1 & 1 & 1 & 1 & 0 & 1 & 1 & 1 & 1 & 1 & 1 & 1 & 1 & 1 & 1 \\
 1 & 1 & 1 & 1 & 1 & 1 & 0 & 0 & 0 & 1 & 1 & 1 & 1 & 1 & 1 & 1 \\
 1 & 1 & 1 & 1 & 0 & 1 & 1 & 1 & 1 & 1 & 1 & 0 & 0 & 0 & 1 & 1 \\
 1 & 1 & 1 & 0 & 1 & 1 & 1 & 1 & 1 & 1 & 1 & 1 & 1 & 1 & 0 & 0 \\
 1 & 1 & 1 & 1 & 0 & 1 & 1 & 1 & 1 & 1 & 1 & 0 & 0 & 0 & 1 & 1 \\
 1 & 1 & 1 & 0 & 1 & 1 & 1 & 1 & 1 & 1 & 1 & 1 & 1 & 1 & 0 & 0 \\
 1 & 0 & 0 & 1 & 1 & 1 & 1 & 1 & 1 & 1 & 1 & 1 & 1 & 1 & 1 & 1 \\
 1 & 1 & 1 & 0 & 1 & 1 & 1 & 1 & 1 & 1 & 1 & 1 & 1 & 1 & 0 & 0 \\
 1 & 0 & 0 & 1 & 1 & 1 & 1 & 1 & 1 & 1 & 1 & 1 & 1 & 1 & 1 & 1 \\
\end{array}
\right)$$
$$X=\left(
\begin{array}{cccccccccccccccc}
 1 & 1 & 0 & 0 & 0 & 1 & 1 & 1 & 1 & 1 & 1 & 1 & 1 & 1 & 1 & 1 \\
 0 & 0 & 1 & 1 & 1 & 1 & 1 & 1 & 1 & 1 & 1 & 1 & 1 & 1 & 1 & 1 \\
 0 & 0 & 1 & 1 & 1 & 1 & 1 & 1 & 1 & 1 & 1 & 1 & 1 & 1 & 1 & 1 \\
 0 & 0 & 1 & 1 & 1 & 1 & 1 & 1 & 1 & 1 & 1 & 1 & 1 & 1 & 1 & 1 \\
 1 & 1 & 0 & 0 & 0 & 1 & 1 & 1 & 1 & 1 & 1 & 1 & 1 & 1 & 1 & 1 \\
 1 & 1 & 1 & 1 & 1 & 1 & 1 & 0 & 1 & 1 & 1 & 1 & 1 & 0 & 1 & 0 \\
 1 & 1 & 1 & 1 & 1 & 1 & 1 & 1 & 0 & 1 & 0 & 1 & 0 & 1 & 0 & 1 \\
 1 & 1 & 1 & 1 & 1 & 1 & 1 & 1 & 1 & 0 & 1 & 0 & 1 & 1 & 1 & 1 \\
 1 & 1 & 1 & 1 & 1 & 1 & 1 & 1 & 1 & 0 & 1 & 0 & 1 & 1 & 1 & 1 \\
 1 & 1 & 1 & 1 & 1 & 1 & 1 & 1 & 0 & 1 & 0 & 1 & 0 & 1 & 0 & 1 \\
 1 & 1 & 1 & 1 & 1 & 1 & 1 & 1 & 0 & 1 & 0 & 1 & 0 & 1 & 0 & 1 \\
 1 & 1 & 1 & 1 & 1 & 1 & 1 & 0 & 1 & 1 & 1 & 1 & 1 & 0 & 1 & 0 \\
 1 & 1 & 1 & 1 & 1 & 1 & 1 & 0 & 1 & 1 & 1 & 1 & 1 & 0 & 1 & 0 \\
 1 & 1 & 1 & 1 & 1 & 1 & 1 & 0 & 1 & 1 & 1 & 1 & 1 & 0 & 1 & 0 \\
 1 & 1 & 1 & 1 & 1 & 0 & 0 & 1 & 1 & 1 & 1 & 1 & 1 & 1 & 1 & 1 \\
 1 & 1 & 1 & 1 & 1 & 0 & 0 & 1 & 1 & 1 & 1 & 1 & 1 & 1 & 1 & 1 \\
\end{array}
\right)$$
\section{Fourier Components \label{App:fourier components}}
Here we show how to calculate the Fourier components
of the tile densities in a ground state configuration. The discrete
Fourier transform of tile number $l$ at the reciprocal vector $\mathbf{G}=2\pi\tau\left(n\mathbf{\hat{x}}+m\mathbf{\hat{y}}\right)$
is 
\begin{equation}
\psi_{l}(n,m)=\frac{1}{N}\sum_{\mathbf{r}}\rho_{l}(\mathbf{r})e^{-i2\pi\tau nx}e^{-i2\pi\tau my}.\label{eq: fourier}
\end{equation}
Using the definitions of $\beta_{1}$ and $\beta_{2}$, we can write
\begin{equation}
\psi_{l}(n,m)=e^{2\pi in\chi_{1}}e^{2\pi im\chi_{2}}\frac{1}{N}\sum_{\mathbf{r}}\rho_{l}(\mathbf{r})e^{-i2\pi n\beta_{1}}e^{-i2\pi m\beta_{2}}.\label{eq:fourirer after going to beta}
\end{equation}
We define $\Omega_{l}$ as the region associated with tile number
$l$ in the $\beta_{1}-\beta_{2}$ torus, defined in figure (\ref{map}).
As $\tau$ is irrational, the functions $\beta_{1}$ and $\beta_{2}$
cover the region $\Omega_{l}$ densely and uniformly, and the infinite sum in (\ref{eq:fourirer after going to beta})
can be turned into an integral: 
\begin{equation}
\psi_{l}(n,m)=e^{2\pi in\chi_{1}}e^{2\pi im\chi_{2}}\iint_{\Omega_{l}}d\beta_{1}d\beta_{2}e^{-i2\pi n\beta_{1}}e^{-i2\pi m\beta_{2}}.\label{eq:fourierafter going to integral}
\end{equation}
This integral can be performed analytically, and we can now find the Fourier components. As an
example, let us calculate the Fourier transform of tile number 2:
\begin{eqnarray}
\psi_{2}(n,m)  =  e^{2\pi in\chi_{1}}e^{2\pi im\chi_{2}}\iint_{\Omega_{2}}d\beta_{1}d\beta_{2}e^{-i2\pi n\beta_{1}}e^{-i2\pi m\beta_{2}}\label{eq:calculating fourier integral}\\
  =  e^{2\pi in\chi_{1}}e^{2\pi im\chi_{2}}\int_{\tau}^{1}d\beta_{1}\int_{1-\tau\beta_{1}}^{\tau}d\beta_{2}e^{-i2\pi n\beta_{1}}e^{-i2\pi m\beta_{2}}\nonumber \\
  =  -e^{2\pi in\chi_{1}}e^{2\pi im\chi_{2}}\frac{im\sin2\pi n\tau+in\tau e^{2\pi i\tau\left(n+\frac{m}{2}\right)}\sin\pi m\tau}{2mn\pi^{2}(m-\tau n)}.\nonumber 
\end{eqnarray}
Using Parseval's theorem, one can now check that indeed the components
corresponding to the reciprocal vectors $\mathbf{G}=2\pi\tau\left(n\mathbf{\hat{x}}+m\mathbf{\hat{y}}\right)$
are the only non-vanishing components. The diffraction pattern is
therefore composed of delta peaks, which confirms the quasicrystalline
nature of each perfect tiling. \\
Having found a general way to calculate the Fourier components of
the ground state, we now define the phases $\gamma_{m}^{x}$ and
$\gamma_{m}^{y}$ (used for the definition of the order parameters $Q_{x}$
and $Q_{y}$ in the main text). We chose the phases such that contributions
of all tile-types to the Bragg peak amplitudes will add coherently in a ground state.
The Fourier components at $\mathbf{G=}2\pi\tau\mathbf{\hat{x}}$ can
be found using equation (\ref{eq:fourierafter going to integral}).
Writing them in the form $\psi_{m}(1,0)=\left|\psi_{m}(1,0)\right|e^{i\zeta_{m}}$,
it is easy to see that the order parameter will have largest possible
length if the phases $\gamma_{m}$ are chosen to be $-\zeta_{m}$
. The same procedure can be used to find the phases corresponding
to $\mathbf{G=}2\pi\tau\mathbf{\hat{y}}$ . The phases $\zeta_{m}^{x}$ and $\zeta_{m}^{y}$ correspond to the center of mass of the region $\Omega_{m}$ on the $\beta_{1}-\beta_{2}$ torus. Table (\ref{tab:A-table-of-gamma})
gives the phases $\gamma_{m}$ of the two order parameters.

\begin{table*}
{\tiny{}}%
\begin{tabular}{|c|c|c|c|c|c|c|c|c|}
\hline 
$m$ & 1 & 2 & 3 & 4 & 5 & 6 & 7 & 8\tabularnewline
\hline 
$\gamma_{m}^{y}/_{2\pi}$ & 0.8090 & 0.8756 & 0.8756 & 0.7424 & 0.7424 & 0.5400 & 0.3368 & 0.2812\tabularnewline
\cline{1-1} 
$\gamma_{m}^{x}/_{2\pi}$ & 0.8090 & 0.5400 & 0.3368 & 0.2812 & 0.0780 & 0.8756 & 0.8756 & 0.7424\tabularnewline
\hline 
$m$ & 9 & 10 & 11 & 12 & 13 & 14 & 15 & 16\tabularnewline
\hline 
$\gamma_{m}^{y}/_{2\pi}$ & 0.0780 & 0.1403 & 0.0780 & 0.4113 & 0.3090 & 0.5400 & 0.20968 & 0.4778\tabularnewline
\cline{1-1} 
$\gamma_{m}^{x}/_{2\pi}$ & 0.7424 & 0.1403 & 0.4113 & 0.0780 & 0.3090 & 0.2068 & 0.5400 & 0.4778\tabularnewline
\hline 
\end{tabular}{\tiny \par}
\caption{\label{tab:A-table-of-gamma}The phases needed to define the order parameters $Q_{x}$ and $Q_{y}$.}
\end{table*}

\bibliographystyle{prsty}

\end{document}